\documentstyle[twocolumn,aps,psfig]{revtex}

\newcommand{\MeV}{{\rm MeV}}                    
\newcommand{\fm}{{\rm fm}}                      
\newcommand{\k}{{\bf k}}                        

\begin{document}

\title{Probing chiral dynamics by charged-pion correlations}

\author{J\o rgen Randrup	}

\address{Nuclear Science Division, 
Lawrence Berkeley National Laboratory\\
1 Cyclotron Road, Berkeley, California 94720, U.S.A.}

\date{December 5, 2000}

\maketitle

\begin{abstract}
The environment generated 
in the mid-rapidity region of a high-energy nuclear collision
endows the pionic degrees of freedom
with a time-dependent effective mass.
Its specific evolution provides a mechanism for the production
of back-to-back charge-conjugate pairs of soft pions
which may present an observable signal
of the non-equilibrium dynamics of the chiral order parameter.
\end{abstract}

\pacs{PACS numbers:
	25.75.-q,
	11.30.Rd,
	05.60.Gg,
	11.10.-z 
}
\narrowtext


High-energy nuclear collisions are expected to produce transient systems
within which chiral symmetry is approximately restored
and the matter is partially deconfined.
The identification and exploration of such a novel phase of matter
is currently a major experimental goal
and the efforts have intensified with the recent commisioning 
of the Relativistic Heavy Ion Collider at BNL.

Through the past decade,
it has been speculated that the rapid expansion of the collision zone,
after an approximate restoration of chiral symmetry has occurred,
may produce long-wavelength isospin-polarized agitations of the pionic field,
commonly referred to as disoriented chiral condensates (DCC),
which in turn should lead to anomalies
in the resulting pion multiplicity distribution.
Reviews of this topic are given in 
Refs.~\cite{Rajagopal:review,Blaizot:review,Bjorken:review}.

Initially the focus was on the expected broadening in the distribution
of the neutral pion fraction
but experimental efforts mounted to search for such a signal
yielded a null result \cite{MiniMax,WA98}.
It has meanwhile become increasingly clear from model calculations
that any DCC effect is carried exclusively by relatively soft pions
and thus the signal can be significantly enhanced
if the analysis is limited to those pions.
Unfortunately,
the neutral pions decay into two photons
and it is practically impossible to perform a reconstruction
that would permit the determination of the individual speeds.
As a consequence,
the neutral pion fraction is less than ideal as an experimental indicator
and the efforts to seach for DCC phenomena 
appear to have arrived at an impasse.

However,
more detailed dynamical calculations suggest that there may be other
signals that would better lend themselves to experimental detection.
In particular,
simulations within the semi-classical linear $\sigma$ model \cite{JR:PRC61}
for cylindrical sources endowed with a longitudinal Bjorken scaling expansion
identified a number of simpler candidate observables.
One is the transverse spectral profile
which is expected to exhibit a marked enhancement below $\approx$$300~\MeV/c$
due to the parametric amplification by the time modulation of the mass.
However,
an experimental extraction of such an effect is hampered by 
the contributions from a number of additional sources of soft pions.
Thus, any observed signal may be less specific
and would need to be correlated with other signatures.
A more promising candidate observable identified in Ref.~\cite{JR:PRC61}
is the fluctuation in the multiplicity of soft pions
which was found to exhibit a significant anomalous increase
with the order of the correlation.
Since a subsequent study \cite{Bleicher} has showed that
such an effect is absent in both RQMD and HIJING
(both widely used event generators using conventional physics input),
it it appears that this observable has a larger specificity
and thus it may be worthwhile to pursue it in the data analysis.

To the above possibilities for DCC signals,
the present communication adds a novel suggestion
for an observable that may be particularly suitable
as an indicator of the phenomenon.
It is based on the key feature that the basic DCC production mechanism
creates neither momentum nor charge
and so a signature may exist in the form of
a large-angle correlation between oppositely charged soft pions.
We first describe the basic effect,
then illustrate the signal by suitable model calculations,
and finally discuss the prospects for its observability.


The initial violent collision of the two approaching nuclei
is followed by a rapid expansion,
which at first proceeds in the longitudinal direction
and then gradually builds up transversly as well.
The pionic degrees of freedom then experience an environment
that is changing accordingly.
Generally,
the effect of the environment can be approximately accounted for
by an in-medium effective mass
which depends both on the degree of agitation
and on the chiral order parameter.
Since the system is steadily cooling down 
while the order parameter reverts from its initial small value
to its large vacuum value in a non-equilibrium fashion,
the effective pion mass has then a correspondingly intricate evolution,
displaying an overall decay towards the free mass
overlaid by the effect of the oscillations by the relaxing order parameter.

Considerable insight into this coupled evolution
can be gained from numerical studies with the linear $\sigma$ model
and though the details depend on the specific treatment,
the emerging qualitative picture is fairly robust.
It is therfore of interest to consider what implications
this type of time evolution of the medium-modified mass
may have on the pion observables.
Scenarios with a time-dependent effective mass occur in many areas of physics,
for example in connection with the inflationary universe,
and they can be treated to various degrees of refinement.
For our present purposes it suffices to recognize
a few general properties of such processes.

In order to bring these out most clearly,
we consider first the simple case where the environment,
and hence the effective mass, is spatially uniform,
as is approximately the case in the interior of the collision zone.
It is then obvious that although the time dependence of the mass
may generate considerable agitation,
this agency cannot add any net momentum.
Thus the any pions produced by the mechanism
must be formed pairwise and moving in opposite directions.
Furthermore,
by a similar reasoning,
the time dependence of the mass does not add any change,
so the produced pairs must be oppositely charged.
Thus,
the particles generated by an arbitrary time dependence
in a uniform medium are charge-conjugate back-to-back pairs.
This basic feature may be exploited as a probe of the chiral dynamcis.


In order to illustrate the effect,
we stay first with the spatially uniform scenario
which is the simplest because the Hamiltonian decouples
so that each degenerate pair of oppositely moving modes
can be considered separately.
A quantum-field treatment is then relatively simple \cite{JR:HIP9}.
The eigenmodes for any given value of the effective mass, $\mu$,
can be obtained by a suitable squeezing of the standard free modes.
Since these modes (which we shall denote as quasiparticles)
diagonalize the problem for the particular value of $\mu$
they provide a particularly instructive
basis for the analysis of the evolution.
We use $\hat{A}_\k$ to denote the annihilation operator
for the quasiparticle modes corresponding to the initial value
of the effective mass, $\mu_i$.
The Heisenberg representation of the corresponding
time-dependent quasiparticle mode is then
\begin{equation}\label{A(t)}
\hat{A}^\nu_\k(t)\ =\
	{U}_k(t) \hat{A}^\nu_\k\ 
	+\ {V}_k(t)^* (\hat{A}^{\bar\nu}_{-\k})^\dagger\ .
\end{equation}
Here the subscript $\k$ denotes the momentum
and the superscript $\nu=\pm$ denotes the charge state
(with $\bar{\nu}\equiv-\nu$).
The time dependence of the quasiparticle operators $\hat{A}^\nu_\k(t)$
derive both from the usual Heisenberg evolution
and from the continual redefinition of the quasiparticle basis.
The Bogoliubov coefficients ${U}_k(t)$ and ${V}_k(t)$
are given in terms of the corresponding mode functions
which in turn are determined by the equation of motion for the pion field.
Being dependent only on the mode frequency $\omega_k$,
they are the same for all charge components $\nu$
and for all momenta with the same magnitude $k$.
It should be noted that the time evolution mixes 
particle states of forward-going positive pions
with hole states of backward-going negative pions,
an expression of the key conservation properties.

Let us now for simplicity assume that the system can initially
be described as a thermal ensemble.
The corresponding occupancy of a given initial
quasiparticle mode is then $N_k=1/[\exp(\omega_k/T)+1]$,
where $T$ is the specified temperature.
The expected occupancy of the mode at a later time $t$
readily follows from (\ref{A(t)}),
\begin{eqnarray}	\nonumber
{N}^\nu_\k(t)\ &\equiv&\  \langle \hat{N}^\nu_\k(t) \rangle\
\equiv\ \langle \hat{A}^\nu_\k(t)^\dagger \hat{A}^\nu_\k(t) \rangle\\
&=&\ |{U}_k(t)|^2 N_k\ +\ |{V}_k(t)|^2 (N_k+1)\ .
\end{eqnarray}

With a bit more elementary operator algebra,
it is possible to obtain the following two equivalent expressions
for the quasiparticle correlation coefficient,
\begin{eqnarray}	\nonumber
\sigma^{\nu'\nu}_{\k'\k}(t)\ &\equiv&\
\langle \hat{N}^{\nu'}_{\k'}(t) \hat{N}^\nu_\k(t) \rangle\ -\
{N}^{\nu'}_{\k'}(t) {N}^\nu_\k(t)\\	\label{sigma1}
&=&\ N_k \bar{N}_k\ \delta_{\nu'\nu}\ \delta_{\k'\k}\
 +\ |{U}_k(t)|^2 |{V}_k(t)|^2\\	\nonumber
&\times&\ \left( 2N_k + 1 \right)^2
[	\delta_{\nu'\nu}\ \delta_{\k'\k} 
+ 	\delta_{\nu'\bar\nu}\ \delta_{\k',-\k} ]\\	\label{sigma2}
&=&\ {N}_\k(t) \bar{N}_\k(t)\ \delta_{\nu'\nu}\ \delta_{\k'\k}\\
\nonumber	&+&\ |{U}_k(t)|^2 |{V}_k(t)|^2
	\left( 2N_k + 1 \right)^2
	\delta_{\nu'\bar\nu}\ \delta_{\k',-\k}\ .
\end{eqnarray}
In the first relation,
the term $N\bar N$ is the initial Bose-Einstein autocorrelation 
for the occupancy of the given mode
while the second term is the additional correlation introduced
by the time dependence of the effective mass, $\mu(t)$,
and its specific form reflects the fact that the autocorrelation
is increased in concert with the back-to-back 
correlation between charge-conjugate pairs.
The last relation shows that the resulting autocorrelation
still has the familiar Bose-Einstein form, ${N}(t)\bar{N}(t)$.

\begin{figure}[h]
\centerline{\psfig{figure=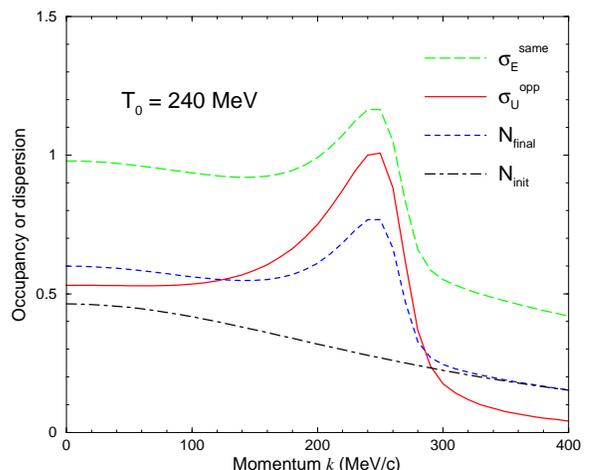,width=3.0in,angle=-90}}
\caption{
The amplification and correlation obtained
for an idealized spatially constant mass function $\mu(t)$
emulating an expanding scenario that starts from an environment
in which the order parameter is $\sigma_0\approx26~\MeV$
(corresponding to $T_0\approx240~\MeV$).
Further explanation is given in the text.
}\label{fig1}
\end{figure}

These features are illustrated in Fig.~\ref{fig1}.
In order to emulate a uniform, longitudinally expanding environment,
we use a simple time dependence of the effective mass,
$\mu^2(t)=m^2+(\mu_i^2-m^2)\exp(-t/t_0)[1+\cos(\omega_{\rm osc}t)]/2$.
With $\mu_i=2m$, $t_0=5~\fm/c$, and $\omega_{\rm osc}=m_\sigma$,
this expression approximates the evolution from an initial environment
in which the order parameter has the value
$\sigma_0\equiv\langle q \bar{q} \rangle\approx26~\MeV$
(corresponding to a temperature of $T_0\approx240~\MeV$ in the 
semi-classical linear $\sigma$ model \cite{JR:PRC61}).

Figure \ref{fig1} shows how
the initial Bose-Einstein quasi\-particle occupation numbers 
$N_{\rm init}$
(dot-dashed curve)
are being enhanced at momenta below $\approx$$300~\MeV/c$ (short dashes).
The resulting peak structure in $N_{\rm final}$
is a reflection of the regularity of the order parameter oscillations:
particular amplification is experienced by the modes that have
their frequency $\omega_k$ in the neighborhood of half the $\sigma$ mass,
corresponding to $k\approx260~\MeV$ for $m_\sigma=600~\MeV$.
The autocorrelation in the mode occupancies (long dashes)
exhibits a similar structure (since it is given by $N\bar N$).
It is noteworthy that although 
the correlation between oppositely moving charge-conjugate pion pairs
(solid curve) is generally smaller than the autocorrelation,
as Eq.~(\ref{sigma1}) dictates,
it acquires a comparable magnitude in the region of strongest amplification
(where the dynamically produced pions dominate over those originally present).
On the other hand, this correlation quickly subsides
as the momentum moves above $\approx$300~MeV/$c$,
bringing out the fact that the effect is confined to the soft regime.
 
The above discussion applies to the idealized scenario
of an entirely uniform environment.
It would be more realistic to consider a mass function
that deviates from the free value only within a finite volume
representing the agitated region.
When the environment has a spatial dependence,
the pions experience forces
which tend to erode the clear back-to-back correlation pattern.
Thus it is important to ascertain the importance of this effect.

\begin{figure}[h]
\centerline{\psfig{figure=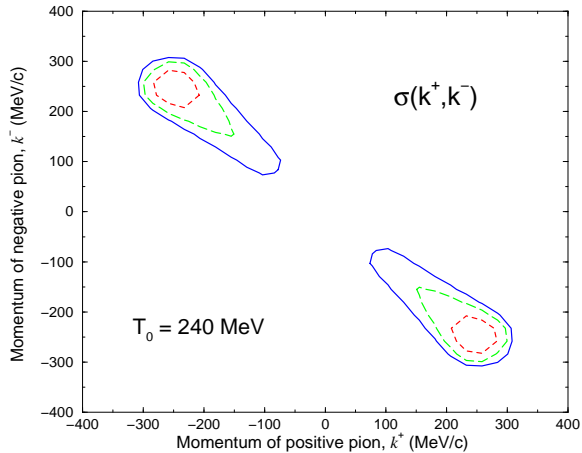,width=3.0in,angle=-90}}
\caption{
Contours of the correlation coefficient for pion pairs of opposite charge
as a function of the final momenta,
for the one-dimensional scenario considered in Ref.~\protect\cite{JR:PRC62}
(which uses a space and time dependent effective mass emulating the
results obtained in Ref.~\protect\cite{JR:PRC61}
for the expansion of a system with 
the initial temperature of $T_0=240~\MeV$ and a initial radius of $R_0=5~\fm$.
The contours are separated by factors of two.
}\label{fig2}
\end{figure}

It was recently shown how the quantum-field treatment 
of Ref.~\cite{JR:HIP9}
can be extended to non-uniform scenarios as well,
still within the effective mass approximation \cite{JR:PRC62}.
We have employed that approach to obtain the time evolution
of the two-body correlation function and
the result is illustrated in Fig.~\ref{fig2}.
As expected,
the correlation pattern seen in Fig.~\ref{fig1} has become somewhat eroded,
due to the spatial dependence of the mass function,
but the basic structure largely persists.
Thus there is reason to hope that it may be experimentally observable. 


In practice,
when making an observation,
it is not possible to identify a single quantum mode $\k$.
Rather,
a collection is made over a phase-space domain that encompasses
many individual modes.
In that situation, one may define the appropriate coarse-grained
particle number operator,
$\hat{\cal N}^\nu_{\cal K}(t) = \sum_{\k\in{\cal K}} \hat{N}^\nu_\k(t)$,
where $\cal K$ denotes a certain domain of $\k$ values.
The associated coarse-grained correlation coefficient
is then of the form
\begin{equation}
\Sigma^{\nu'\nu}_{\cal K'K}\ \equiv\ 
\langle \hat{\cal N}^{\nu'}_{\cal K'}(t) \hat{\cal N}^\nu_{\cal K}(t) \rangle\
-\ 	\langle \hat{\cal N}^{\nu'}_{\cal K'}(t)\rangle
	\langle \hat{\cal N}^\nu_{\cal K}(t)\rangle\ .
\end{equation}
To examine the effect of such a coarse graining,
we let $\cal K$ encompass all positive pions moving forward
with a momentum up to $k_0$
while $\cal K'$ includes all negative pions moving backward
with a momentum down to $-k_0$.

\begin{figure}[h]
\centerline{\psfig{figure=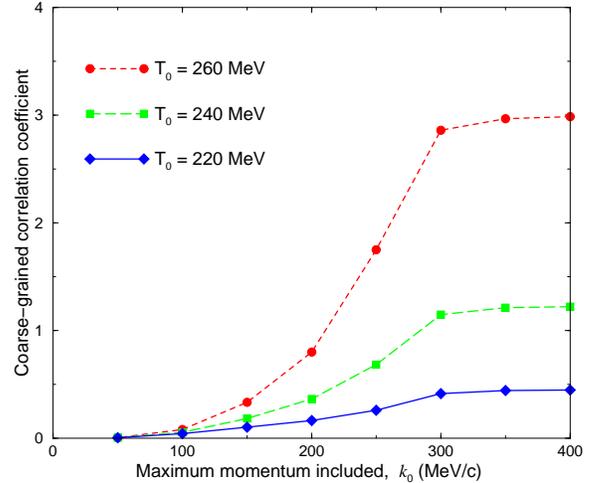,width=3.0in,angle=-90}}
\caption{
The coarse-grained correlation coefficient
for oppositely moving charge-conjugate soft pions
as a function of the cut-off momentum $k_0$,
for three different initial scenarios.
}\label{fig3}
\end{figure}

Figure \ref{fig3} shows 
the resulting coarse-grained correlation coefficient
for three scenarios corresponding 
to different initial degrees of agitation and restoration.
In all cases the coarse-grained correlation coefficient saturates
once the cut-off $k_0$ is above $\approx$$300~\MeV/c$,
the maximum momentum for which the amplification mechanism is effective.
Thus there is no gain in signal by extending the cut-off to higher values,
while the background increases.
In fact, in order to reduce contamination from $\rho$ decays (see below),
it may be desirable to employ a momentum cut-off below $300~\MeV/c$
and the results suggest that this might well be possible.
In the other direction, at the lower momenta,
it is practically difficult to measure pions 
with $k_\perp$ much below $100~\MeV/c$
but these are seen to contribute relatively little to the total signal.

Another important feature brought out in Fig.~\ref{fig3} 
is the good sensitivity of the signal to the physical scenario.
For the relatively moderate range of initial temperatures considered, 
$T_0=220,240,260~\MeV$,
the initial chiral order parametertakes on the values
$\sigma_0=19,26,42~\MeV$
and these differences have become magnified
in the resulting correlation signal.
This feature demonstrates that the signal is in fact quite sensitive
to the degree of chiral restoration achieved early on.
 

The purpose of this note is to draw attention
to the possible utility of analyzing the large-angle correlations
between soft charge-conjugate pion pairs.
The focus has been on arguing that one would expect
an effect on rather general grounds,
based simply on the non-equilibrium evolution 
of the chiral order parameter
which in turn endows the pion with a temporal modulation of its effective mass 
that provides the possibility for parametric amplification.
Moreover,
within the simple linear $\sigma$ model,
we have tried to achieve an idea of the magnitude and persistence
of the expected effect.
Of course,
this ``signal'' is partially obscured or eroded
by a number of other processes 
and thus any attempt to extract it from the experimental data
must take careful account of such ``background'' contributions.

One concern is the possible degradation of the primary signal
in the course of the propagation of the pions 
as they leave the interaction zone.
As the signal-carrying pions are soft they might easily be deflected
significantly by encounters with other hadrons in the later stage
of the reaction process.
However,
the very mechanism by which they are produced,
the oscillatory relaxation of the chiral order parameter,
guarantees that they appear only relatively late.
Simple estimates, as well as more elaborate numerical studies,
show that it takes several to many $\fm/c$ for the amplification mechanism
to complete (say $5~\fm/c$).
At this time,
most of the material in the collision zone has already dispersed,
so the DCC pions will be born in a relatively empty environment
(which deviates from the regular vacuum pricipally by the deviation
of the order parameter from its vacuum value).
Moreover,
since they are soft,
they will have little chance of catching up with the expanding shell
of regular collision debris.
Thus,
one may expect that the pions of interest will in fact propagate
to the detector relatively undisturbed.
These features are rather similar to those of the
``Baked Alaska'' scenario discussed by Bjorken \cite{Bjorken:review}. 

Another important issue concerns the possible presence
of other agencies that may lead to a similar signal
and thus obscure the signal of interest.
While there are many physically different sources 
of charge-conjugate pion pairs,
fortunately only few lead to strong back-to-back correlations.
It is particularly inportant to discuss 
the dominant decay of $\rho(770)$, 
$\rho\to\pi^+\pi^-$.
Although the two pions are back-to-back correlated,
they emerge from a $\rho$ meson at rest with momenta of about $360~\MeV/c$
which is somewhat above the upper limit of the expected effect
($k_{\rm max}\approx300~\MeV/c$).
Moreover,
while the decay of a $\rho$ meson in motion
may well contribute a single pion to the yield below this limit,
the contribution to the coincidence yield is insignificant,
even when taking account of the final $\rho$ width and
the in-medium thermal distortion.
Thus, in the relevant domain of soft pions,
the $\rho$ decays are not expected to obscure to the two-body DCC signal.

It should also be noted that although
$\eta(550)$ and $\omega(780)$ may contribute  $\pi^+\pi^-$ pairs,
these all arise in three-body decays
which renders them only rather weakly correlated
and so they should not pose a serious problem.

In conclusion, then,
we suggest that the data now being taken at RHIC
be analyzed for indications of the
described signature in the large-angle correlation 
of soft charge-conjugate pion pairs.
It may also be worthwhile to scrutinize existing SPS data
for this signal.
If indeed identified,
this signal may offer a means for probing the
degree of chiral restoration achieved and the subsequent DCC dynamics.

As a final comment,
we wish to note that the considered effect
should also manifest itself in electromagnetic observables,
photons and dileptons.
For one thing,
the effect is equally well present in the neutral pions
which generally decay into a photon-pair before detection.
Though it is impractical to reconstruct their pre-decay momenta,
the excess of soft neutral pions should 
make a corresponding contribution to the photon yield.
Moreover
charge-conjugate pion pairs may annihilate to produce 
either photons or dileptons
and the soft excess pions discussed here would then contribute accordingly
to these observables.
Therefore,
should the analysis of the charged-pion data suggest 
the presence of a soft-pion excess,
this finding must be taken into account
in the analysis of the electromagnetic signals.

\section*{Acknowledgements}
This work was supported by the Director, Office of Energy Research,
Office of High Energy and Nuclear Physics,
Nuclear Physics Division of the U.S. Department of Energy
under Contract No.\ DE-AC03-76SF00098
and by the Gesellschaft f{\"u}r Schwerionenforschung,
Darmstadt, Germany.
The author also wishes to acknowledge helpful comments by
B.~Friman, J.~Stachel, and R.~Vogt.



\vfill{\small \noindent{\em LBNL-47168 (2000): \hfill Physical Review C}}

			\end{document}